# A special cross-tie domain wall in helimagnet

Yuan Yao[1,4*], Bei Ding[1,4], Jun Liu[1], Jinjing Liang[1,3], Hang Li[1], Xi Shen[1], Richeng Yu[1], Wenhong Wang[1,2,3*]

1 Beijing National Laboratory for Condensed Matter Physics, Institute of Physics, Chinese Academy of Sciences, Beijing 100190, China

2 Songshan Lake Materials Laboratory, Dongguan, Guangdong 523808, China

3 University of Chinese Academy of Sciences, Beijing 100049, China

4 These authors contributed equally: Yuan Yao and Bei Ding

Corresponding Author
*E-mail: yaoyuan@iphy.ac.cn
　　　　wenhong.wang@iphy.ac.cn

**Abstract**

A special cross-tie (SCT) domain wall was discovered in the helimagnet MnCoSi alloy via the magnetic vector field tomography in Lorentz transmission electron microscopy (LTEM). Different to the traditional cross-tie (TCT) domain wall where the convergent/divergent magnetic moment configuration line up one by one, the relative large Bloch type sub-walls emerge in this brand-new SCT domain wall and two mutually perpendicular rotation axes coexist in this special feature. The straight magnetic stripes accompanied with the unraveled domain walls hint the complex mechanism to form this SCT structure. Interestingly, different orientation of this domain wall in LTEM can easily exhibit various magnetic features, including meron/antimeron chains or bimeron strings.

**Introduction**

Helimagnets exhibit diverse micro-magnetic structures including those non-trivial topological features, such as solitons [1], skyrmions [2], merons [3] or hedgehogs [4], because of the competition between their non-collinear or non-coplanar magnetic moments with anisotropy, defects/defaults or even shape boundaries of the materials. These novel structures have received lots of attentions owing to their potential application in future spintronic devices. Investigating the accurate magnetic features is an essential demand to understand the principles dominating the formation and dynamic behavior of these special micro-magnetic structures. Among various approaches employed to characterize the configurations of magnetic structures, Lorentz transmission electron microscope/microscopy (LTEM) demonstrates its powerful ability in directly disclosing and identifying the characteristics of the magnetic objects with high spatial resolution [5]. Direct imaging [6], holography [7] and renascent differential phase contrast (DPC) [8, 9] are the main LTEM streams to probe the small magnetic patterns in materials. The serials focus imaging helped by transport of intensity equation (TIE) [10-12] method is a convenient technique due to its simple operation and brief data processing. However, the principle of LTEM determines that only the projection of the magnetic structure is displayed on image so that the investigation for the real features is restricted because of missing information along the illuminating electron beam. To retrieve the configurations of magnetic moment in three-dimensional (3D) space, many images should be acquired at different tilting angles and vector field tomography algorithm is used to map the 3D magnetic inductions of the specimen in entire space. This pipeline successfully reconstructed the accurate features of the Dzyaloshinskii-Moriya interaction (DMI) [13] and dipole-dipole interaction (DDI) skyrmions [14] and testified that the surface moment configurations of these microscopic magnets are the convergent or divergent Neel type features, reminding a caution for interpreting the magnetic patterns investigated by those metrologies which gauge the surface magnetic vectors sensitively.

Among those magnetic fancies, the merons discovered with sophisticated approaches in different magnetic systems display versatile characteristics in the literature, such as meron lattice[3], meron chains[15], meron pairs [16, 17] and bimerons[18]. The swirling style of the various merons is determined by their intrinsic topologic number involving the charity and polarity, however, the characterization methods indeed seldom investigate the actual configuration of those micro-structures in experiments. In addition, the presented two dimensional images of the real magnetic structure can be easily affected by the work principle, spatial resolution, processing parameters and even post-processing tricks. For example, the key discrimination between typical skyrmions and merons is that the moments on periphery of skyrmions possess the out-plane components but those of merons still keep the complete in-plane rotation [3]. However, sometimes it is difficult to distinguish that tiny difference if the edge of the magnetic structure is close to the spatial resolution limit of the characterization capability. Considering that most investigations only detect particular components of the magnetic field locating on the surface layer or overall projecting along special direction, and heavily rely on data processing to retrieve the signals, the artificial or bias could be introduced easily and misguide the interpretation of the experimental results. Fig. 1a and b shows a traditional cross-tie (TCT) domain wall with its schematic diagram and experimental LTEM features. The retrieved in-plane configuration projected along the illumination electron beam exhibits the convergent and divergent sequence within the domain wall which separates two antiparallel uniform domains while the cores of the vortex/anti-vortex should contain the out-plane components, according to the schematic model. If just focusing on the local substructures TCT domain wall could be recognized as the meron/antimeron chain or bimerons.

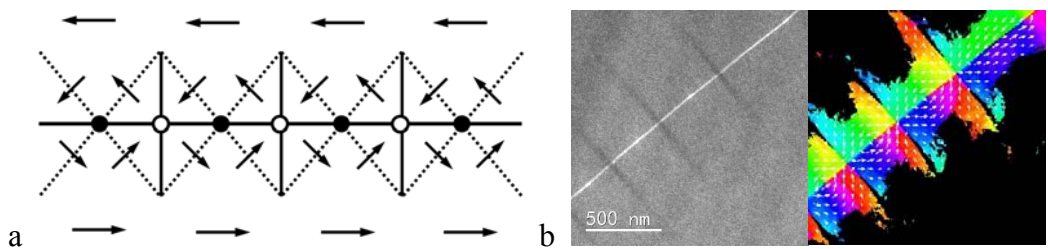

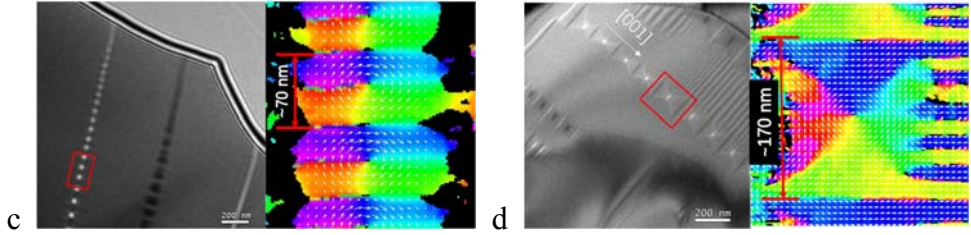

Fig. 1 a) The schematic diagram of a TCT domain wall, b) LTEM image of a TCT domain wall and the retrieved in-plane moment orientation, acquired in $Fe_5Sn_3$ sample. c) and d) LTEM images and in-plane direction mappings of the bead strings and sandglass sequences in helimagnet MnCoSi, respectively. The periods of the substructures in c) and d) are about 70 nm and 170 nm.

MnCoSi is a helimagnet with an incommensurate spiral pitch along its c axis [19]. Our previous neutron diffractions and in-situ LTEM analysis disclosed that its spiral configuration changes to cycloidal stripes at low temperature [20]. In this letter, we report a special cross-tie (SCT) domain wall separating two stripe domains in MnCoSi, where the vortices are isolated by Bloch type sub-walls, different to the general TCT domain wall which uses the anti-vortices to segment the vortices. Therefore, two mutually perpendicular rotation axes of the moments coexist within the new magnetic feature, unlike in TCT domain wall.

**Experiments**

Tetragonal MnCoSi samples for LTEM characterization were prepared with usual ion-milling for polycrystalline bulk and focused ion beam (FIB) milling for single crystal flake, respectively. The facet of FIB specimen was (110) plane and utilized for magnetic moments reconstruction. The incommensurate helical vector along [001] hard axis, confirmed by both neutron powder diffraction and LTEM characterization [20]. During the exploration of the helical magnetic stripes perpendicular to [001] direction and their dynamic behavior at different temperature, we also find some orderly bright/dark strings in the thinner region of the specimen (Fig. 1c and d). These strings exhibit various appearance, bead or sandglass sequence with different intervals, sometimes surrounded by the straight stripes. The direction of the strings is along the [001] axis of MnCoSi, determined from the stripe feature. The in-plane magnetic moments retrieved by TIE algorithm show that the beads are the vortex features separated by the divergence structures, which is similar to the merons chains or bimerons in the literature but the sandglass features are the crowded vortices with the sharp straight interface. It seems that there are different types of magnetic micro-structures in MnCoSi besides the intrinsic incommensurate magnetic stripes.

Our previous studies in skyrmions indicate that the orientation of the magnetic configurations can lead diverse pictures due to the projection characteristic of LTEM imaging principle [12, 21]. Thus images only taken along one orientation are not sufficient to conclude the real magnetic structures. To clarify the accurate magnetic structures observed in MnCoSi, we employed vector field tomography [22-24] to reconstruct the 3D magnetic induction distribution. A bead chain with spaces from ~85 to 122 nm (Fig. 2a) was selected and tilted around two orthogonal axes (x or y direction shown in Fig. 2a) and at each tilting angle, over-focused, in-focused and under-focused images have been acquired to calculate the corresponding in-plane Bx (for x-tilting) or By (for y-tilting) mappings. The tilting range was from -63° to 60° and the interval smaller than 5° was not constant during tilting to avoid the destructive dynamic diffraction contrast at some particular crystal orientation. The calculated Bx and By stacks for different angles were input into the reconstruct program (wolf) to produce the 3D Bx and By matrixes, respectively. At last the 3D Bz matrix can be deduced from the Bx and By datasets according to the relation $\vec{\nabla} \cdot \vec{B} = 0$ [23, 24].

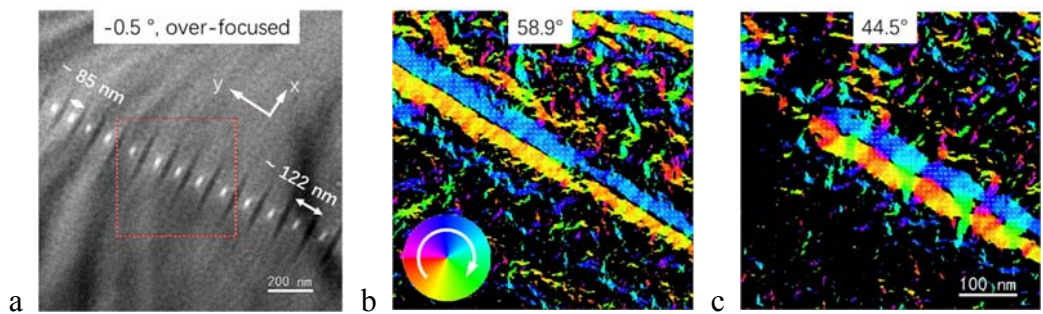

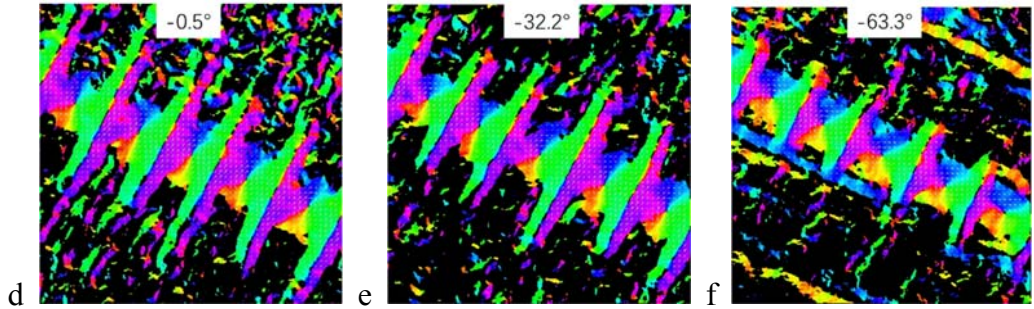

Fig. 2 a) Un-tilted image to show the feature of the chain, where the red square is the investigated region for following mappings. (b-f) Orientation mappings of in-plane moments for different y-tilting angle. The appearance of the configuration varies obviously.

The showcase of retrieved orientation mappings of in-plane moment at different y-tilting angles depicts that the sharp boundaries between the vortices shorten gradually as the specimen tilting so the magnetic features change to the bimeron chain (around 44.5° ) and even transit into two antiparallel stripes at high angle (around 60°) but at -63.3° the initial state still remains. Fig. 3 outlines the recovered 3D magnetic inductions with streamlines, colorized by the orientation of x-y plane moments. The solenoidal moments swirl strongly in the vortices with the diameter below 100 nm. Few of [001] components appear within the longer boundaries segregating the vortices means that the interfaces between the vortices are the Bloch wall where the moments quickly flip 180° within 10 nm. The y-view (x-z plane in Fig. 3) demonstrates the sloping spiral axis and explains the asymmetric variation of the features during sample tilting – positive tilting reaches the side-view of the vortices quickly than negative tilting. The view in y-z plane also illustrates the string direction [001] deviated from the rotation axis (y axis).

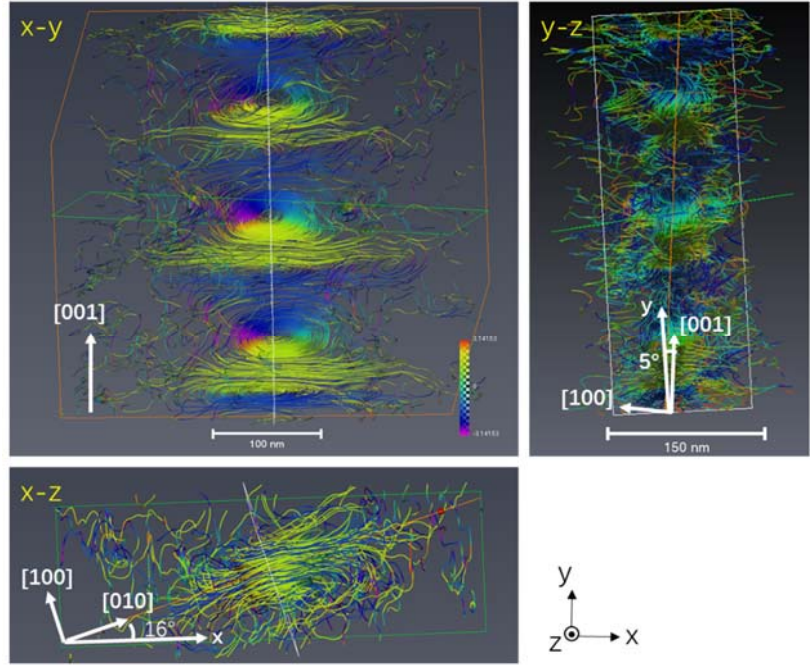

Fig.3 The streamlines of the reconstructed magnetic vector field rendered with orientation of the x-y plane components.

Tomography result verifies that the equatorial planes of the vortices are (100) facets and each vortex twirls around [100] direction with same charity. The interfaces between the vortices are the (001) facets where the revolving moments lie. Thus, two mutually perpendicular rotation axes, one is of the vortex along [100] and other is of Bloch sub-wall along [001], survive together in the investigated structures. It is a novel magnetic configuration different to the TCT or meron/antimeron model where the vortices/anti-vortices line themselves side by side.

It should be noted that vortex strings appear as the zippers sealing two regions embodying straight stripes in many scenes (Fig. 4a-c). The periodic of the incommensurate helical magnetic structure in MnCoSi is about 5.7 nm at room temperature, confirmed by electron diffraction patterns and high magnification images [20]. But the broader stripes with pitch of more than 10 nm were often observed in the experimental data. These broader stripes might be the antiparallel [010]-orientating moments transited from the intrinsic helical periods under some uncertain influences, however it still needs more evidences. Experiments confirmed that only when electron beam is precisely along [100] direction these stripes can be distinguished clearly in the images. Larger defocus which is usual in LTEM to enhance the image contrast sometimes blurs these tiny features. Those slender stripes in the thinner specimen regions also fail to contribute adequate contrast in the images. The inclination of the sample, such as the tomography experiments shown in fig. 3, may be another factor leading to the disappearance of the stripes. Whatever the accurate style of the stripes clamping the vortex strings, at least in many situations it is undoubted that the separated parts by the vortex string are not two antiparallel uniform domains.

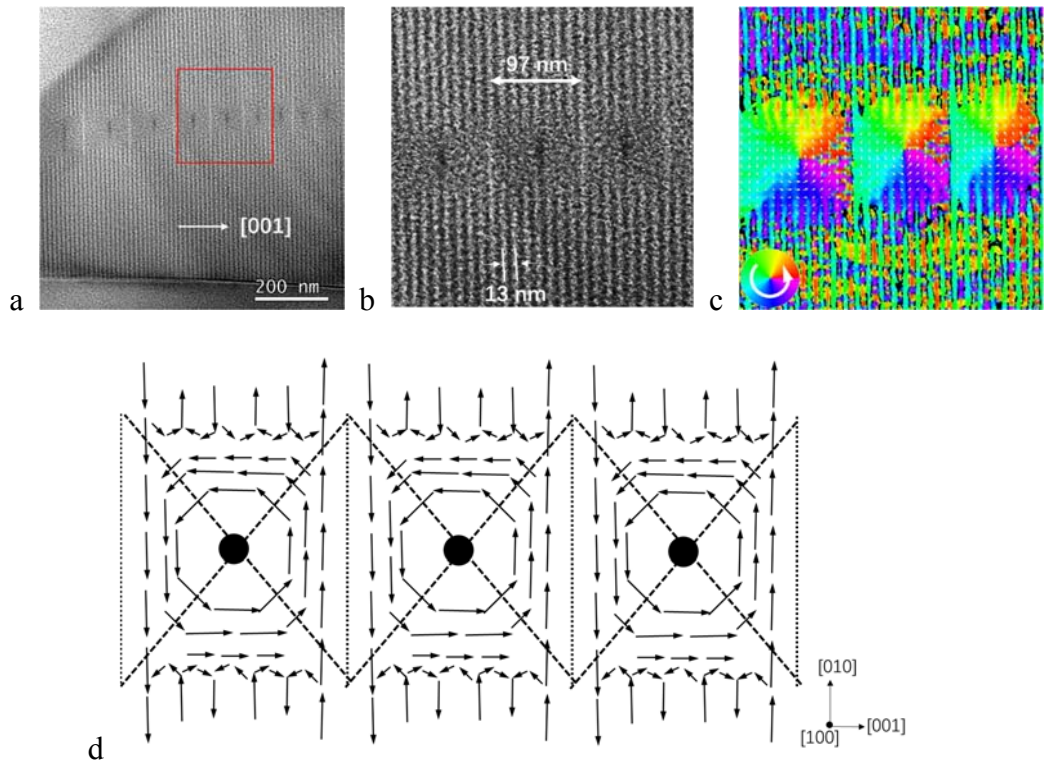

Fig. 4 a)-c) The straight stripes surrounding the domain wall. d) Schematic diagram of the new SCT domain wall sealing the stripe domains.

This new SCT domain wall possesses some unique characters. First, they are strait and always associated with a parallel counter with totally inversed spiral manner. There are signs that a Bloch sub-wall in one string directly reaches another Bloch sub-wall with same moment orientation through a stripe (Fig. s7 in supplemental information). Thus there is a mismatch between two coupled sequences to satisfy this physical demand. Second, the distance of the substructure within the domain wall is not a constant, as well as its partner. Although some hints imply the space is modulated by the thickness of the investigated region, the reason still is an open question now. Third, the strings are precisely along the hard axis [001] of the crystal MnCoSi without anti-vortex substructures to isolate the vortices as suggested in TCT wall, substituted by the larger 180° Bloch sub-wall. Two Bloch sub-walls always span several straight stripes, inferring that the evolution of such domain walls should be more complex beyond simply constructing it with out-of-phase among those fine units. Low magnification images (Fig. s8 in supplemental information) show that some of these special domain walls evolve to large domains at thicker parts of the specimen. TCT domain wall originated from the competition between the dipolar exchange and the static magnetic energy survivals within a critical range of thickness, but the helical magnetic structures in MnCoSi can weaken the effect of static magnetic energy because the intrinsic helical moments balance the possible magnetic charges on each surface of the TEM specimen so that thickness or the static magnetic energy could not contribution to the formation of SCT domain wall. Novel topological domain walls have been revealed in FeGe helimagnet where the DMI plays an important role to reform the boundary shapes between the helical stripe domain [25], similar to the style in our material, but MnCoSi is a centrosymmetric crystal and the study of MnCoSi is unfortunately inadequate now. Many important physical properties, such as the anisotropy, exchange coefficient, the origin of the helical feature, are still waiting for prospection. This restricts the theoretical analysis or prediction of the mechanism behind the experimental appearances, and therefore micro-magnetic simulation confronts a huge obstacle to imitate the observed phenomena. On the other hand, the poor resolution of the LTEM also eliminates more structure inspection nearby the frontier between the vortex and the stripes, especially for 3D reconstruction of the magnetic configurations around this region. The precise structure of the stripes with different periods is still waiting to be classified. Diversiform features of magnetic microstructures demonstrate the complexity to disclose the origination of the new domain wall in the helimagnet MnCoSi alloy.

In summary, a new cross-tie-like domain wall has been discovered in helimagnet MnCoSi with the LTEM tomography approach. The swirling vortices are fenced by Bloch wall substructures. This special domain wall connects two stripe domains and contains two mutual perpendicular spiral axes within the domain wall. At different crystal orientation the projection of the moments varies and displays multifold feature which may confuse the investigators.

**Acknowledgement**

The authors thank Prof. Ying Zhang for the warm support in TEM characterization. This work was supported by the National Natural Science Foundation of China (Grant No. 11874410), the National Key R&D Program of China (2017YFA0303202, 2017YFA0206200) and the Strategic Priority Research Program of the Chinese Academy of Sciences (No. XDB33030200).

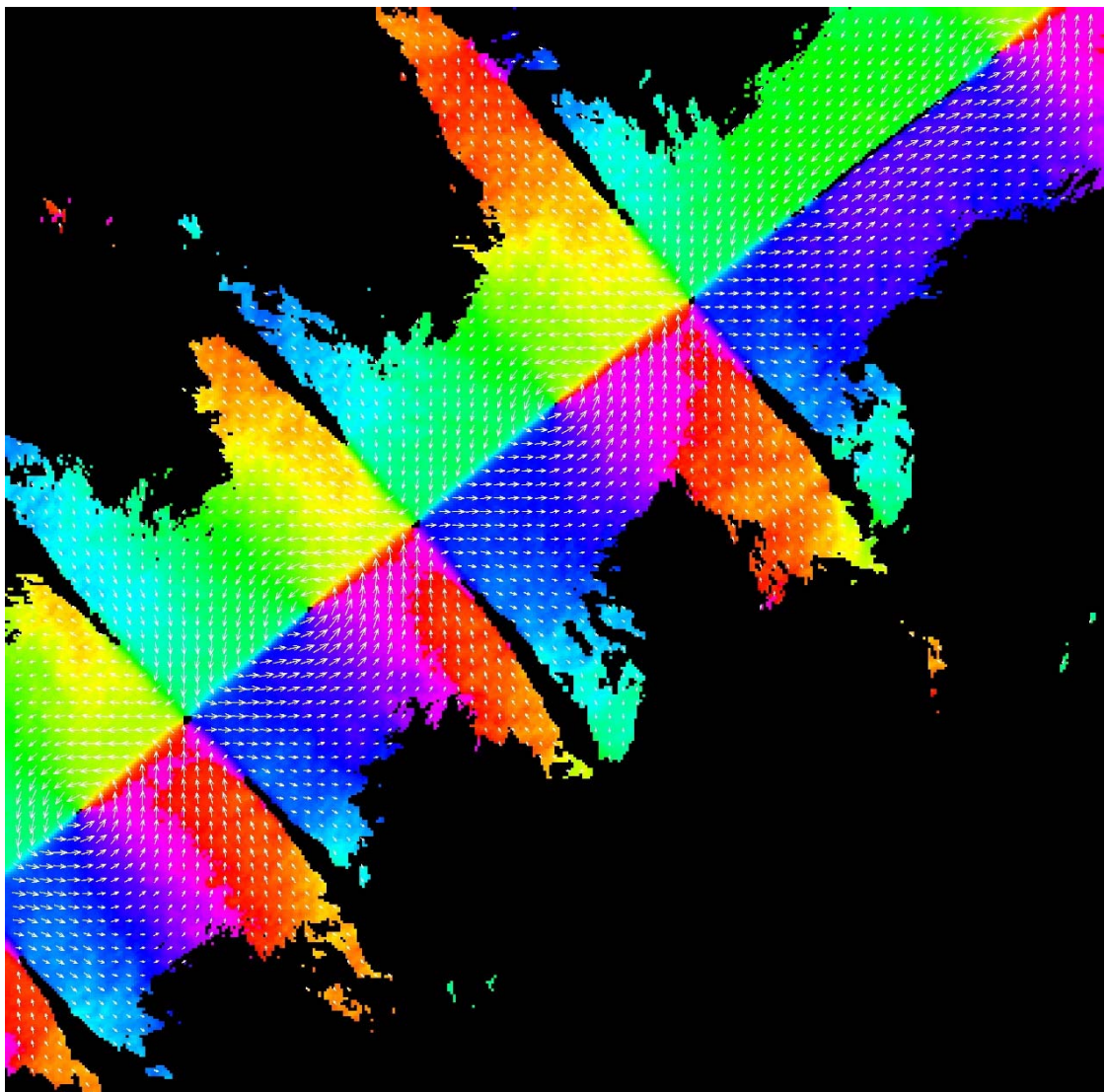
Fig. s1 The in-plane moment orientation of traditional cross-tie domain wall in fig. 1b

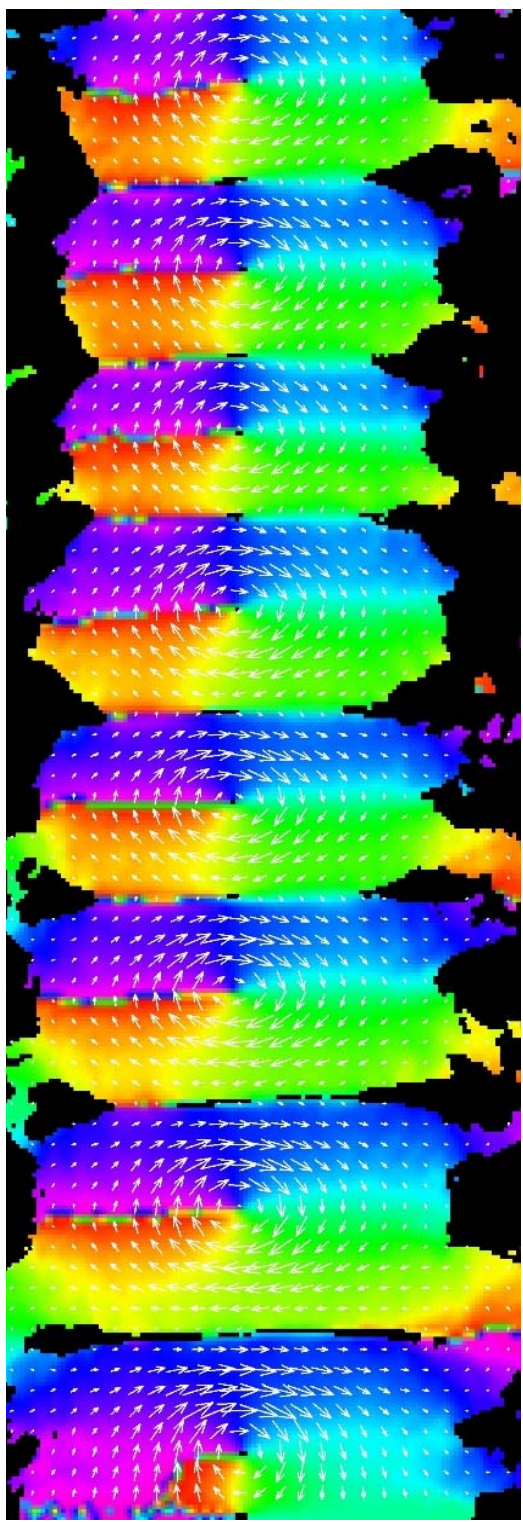

Fig. s2 The in-plane moment orientation mapping in Fig. 1c

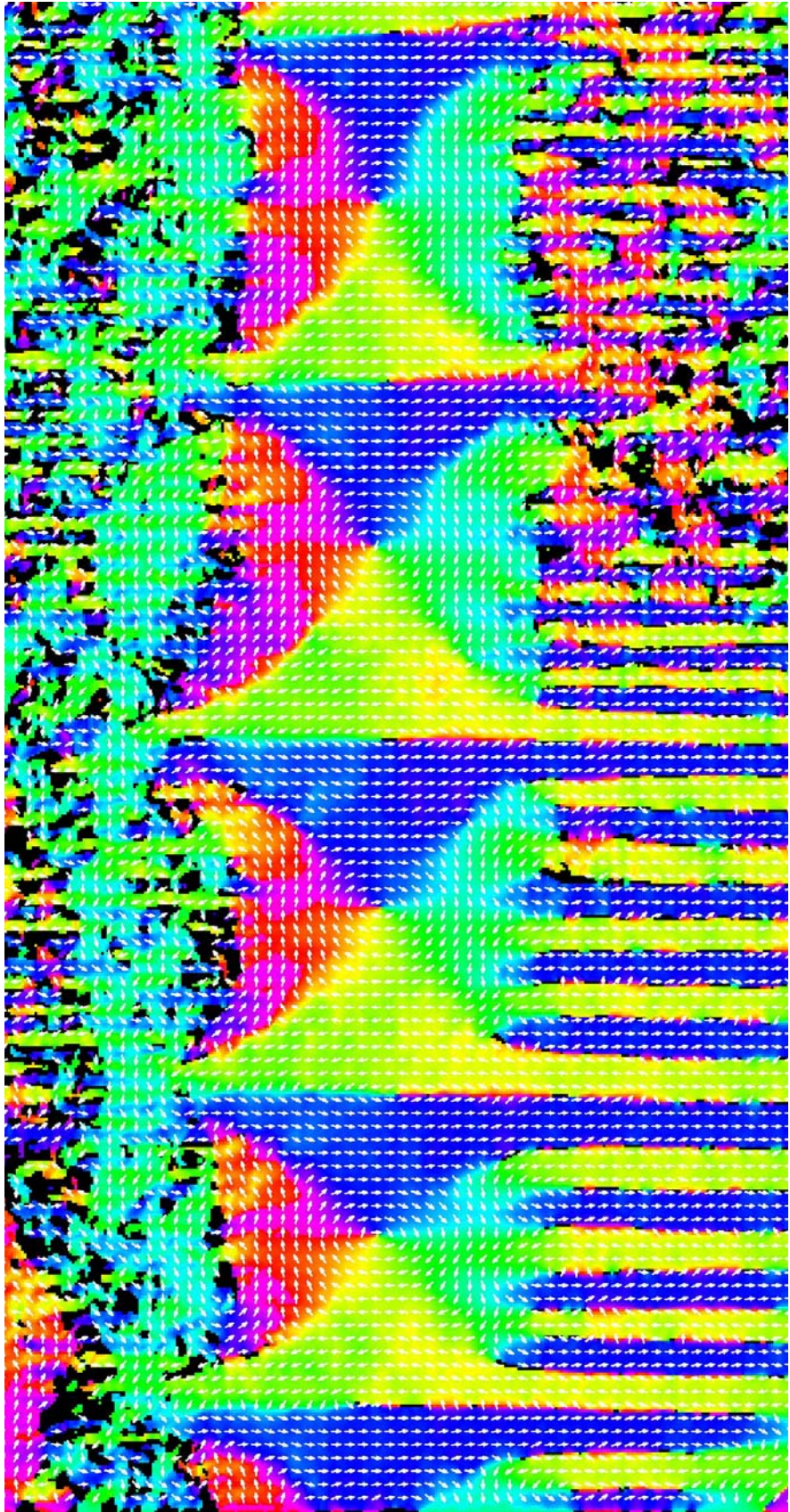
Fig. s3 The in-plane moment orientation mapping in Fig. 1d, with the straight stripes.

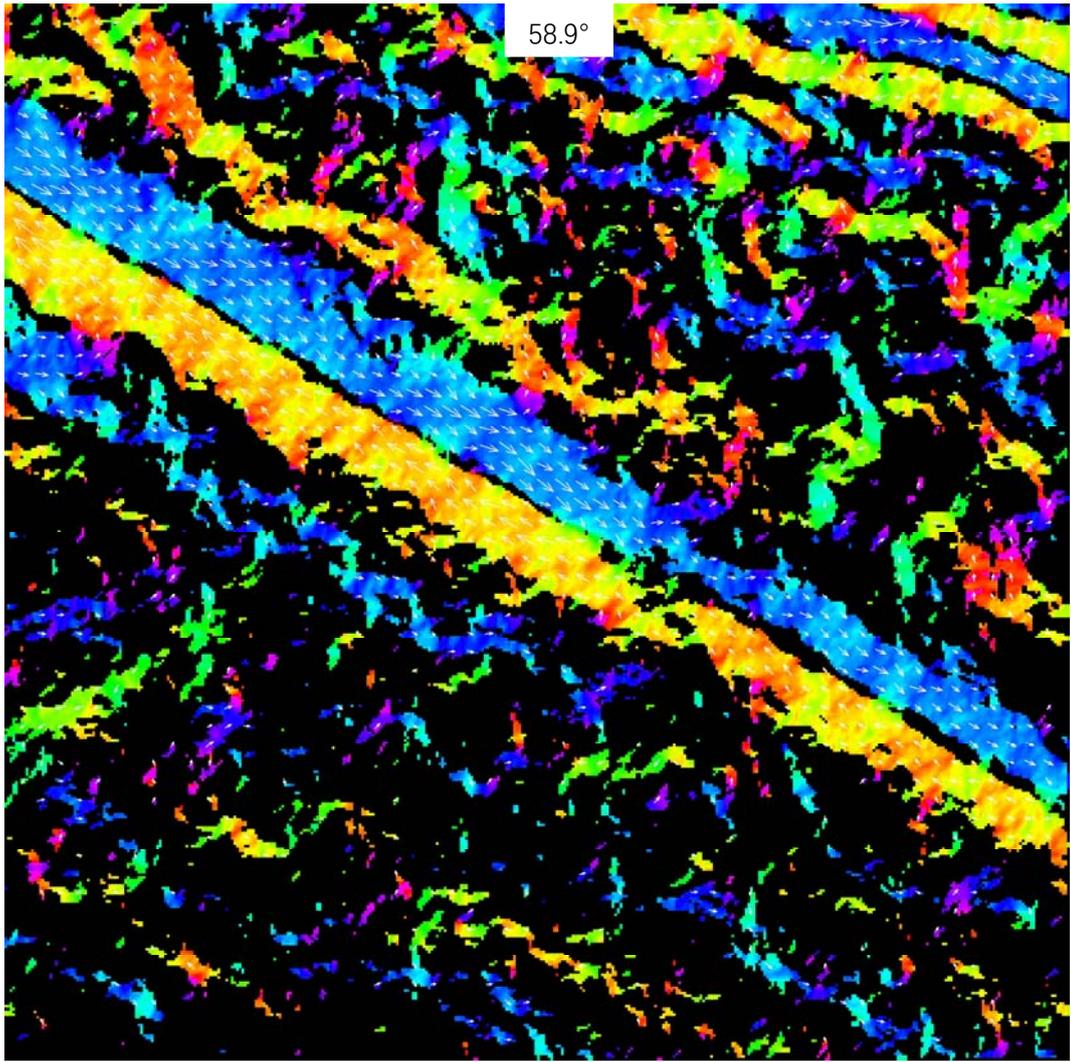

a)

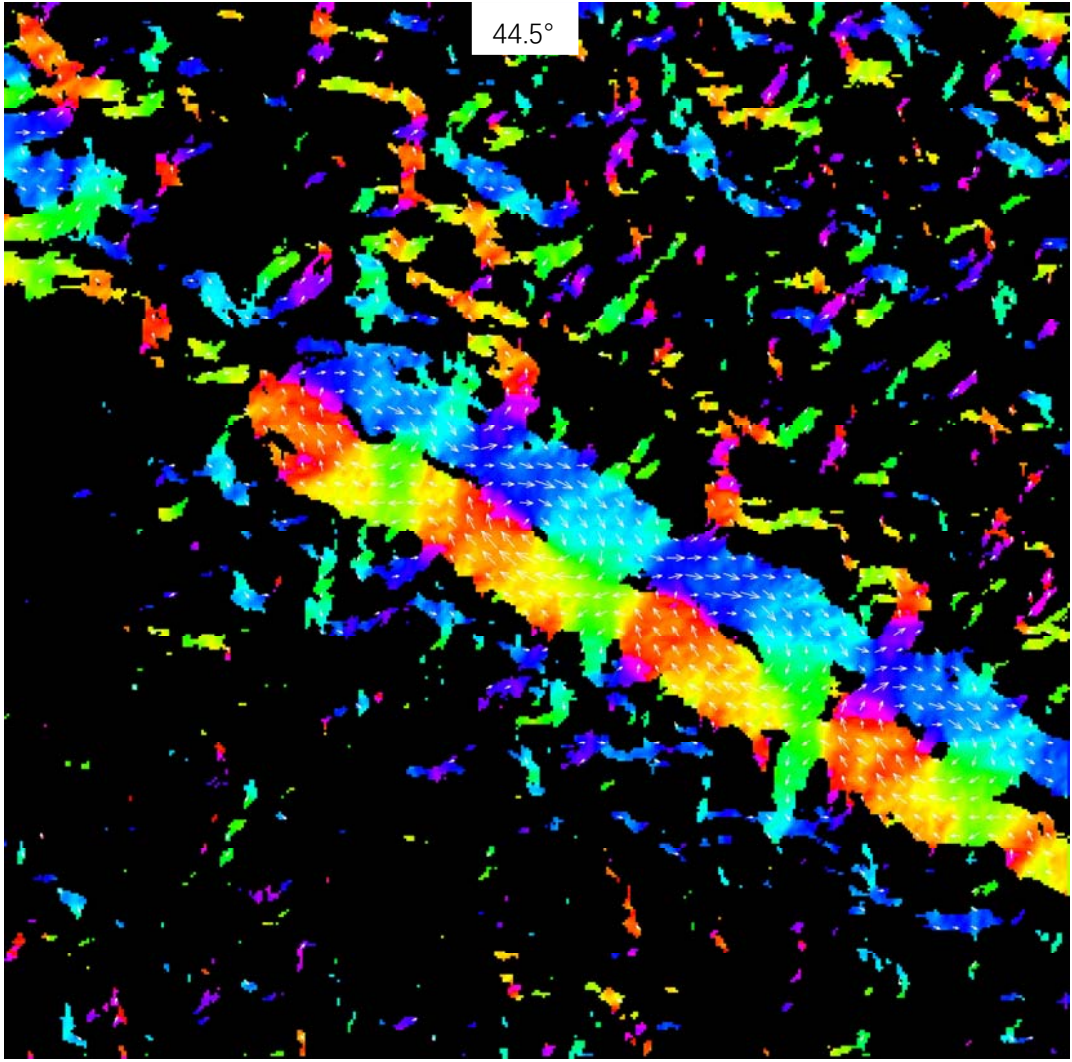

b)

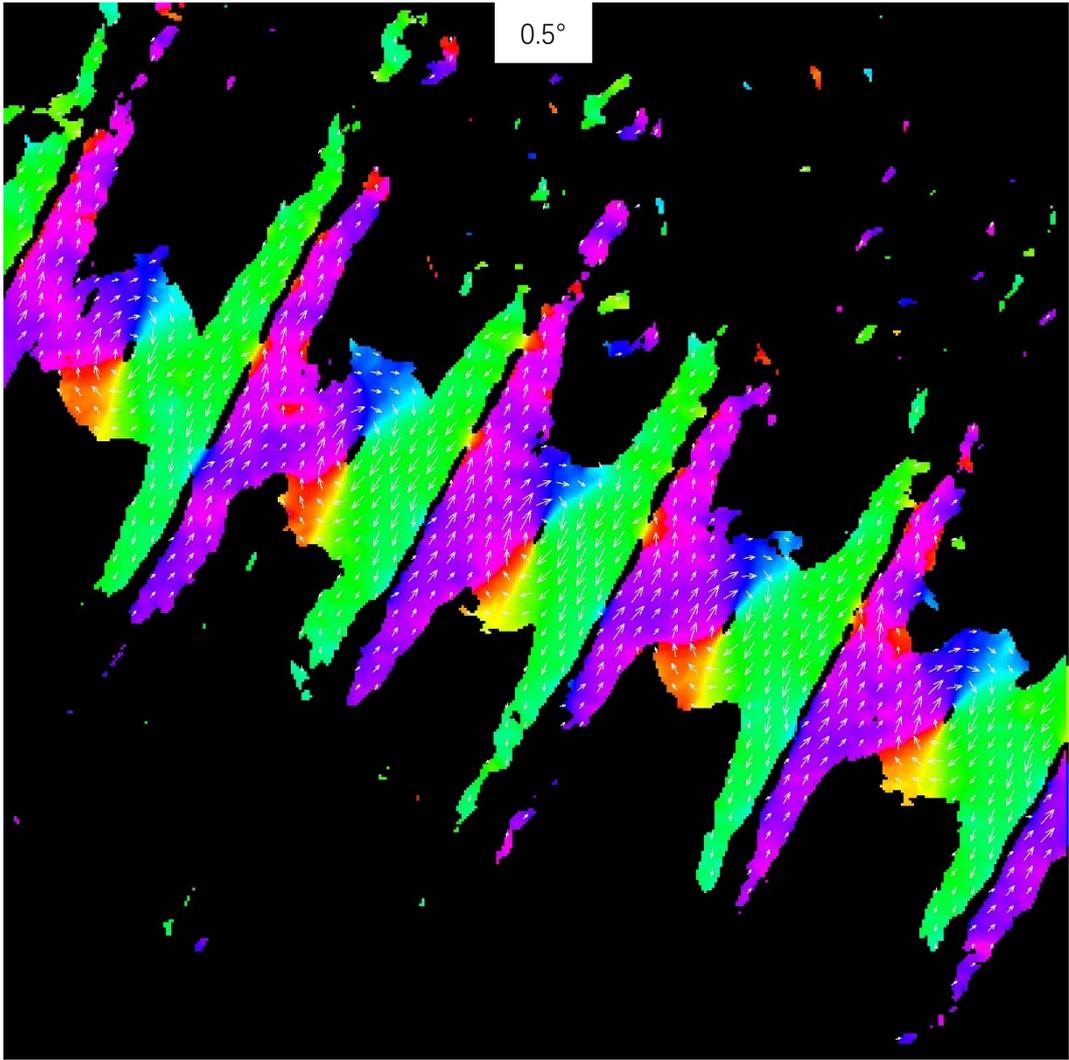

c)

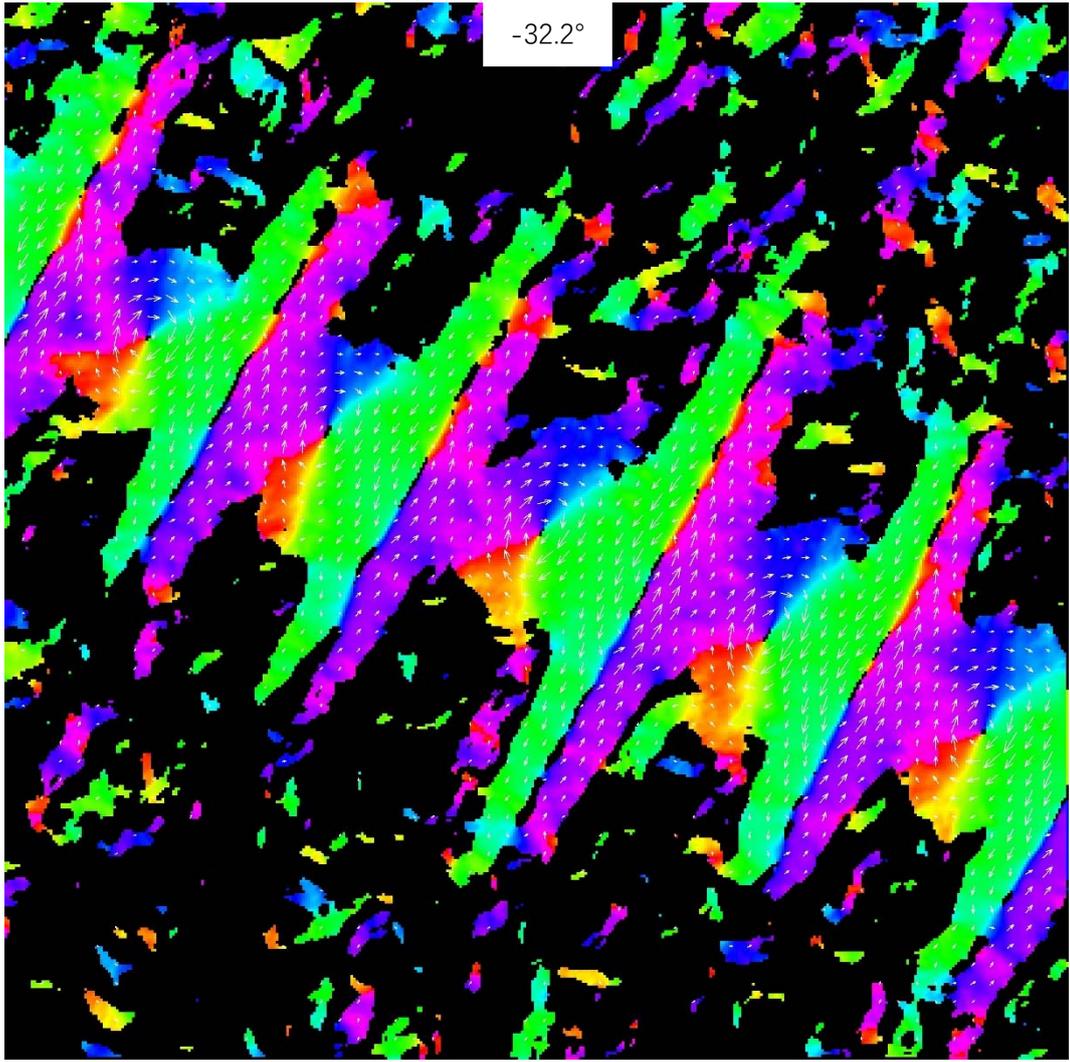

d)

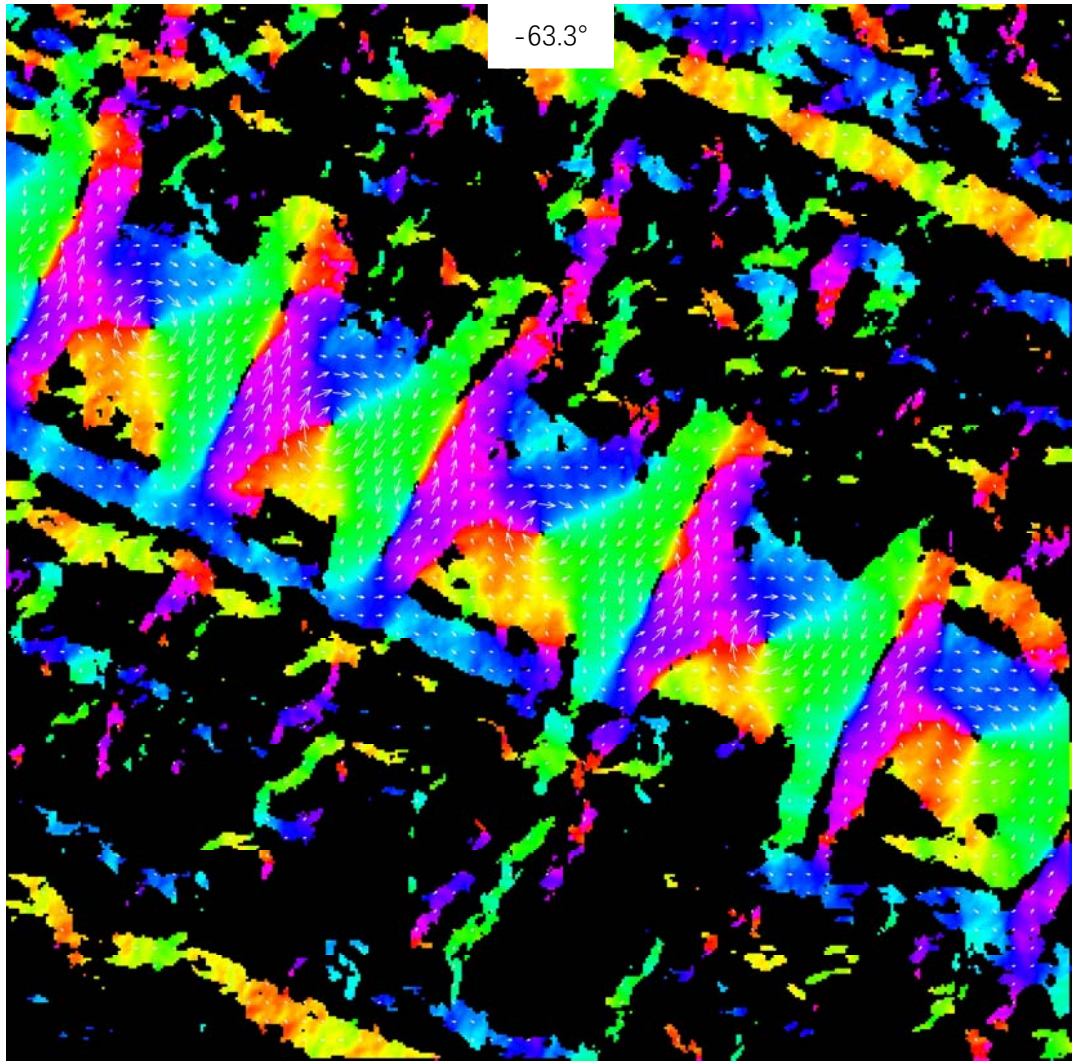
e)

Fig. s4 a-e) The in-plane moment orientation mapping in Fig. 2 (b-f)

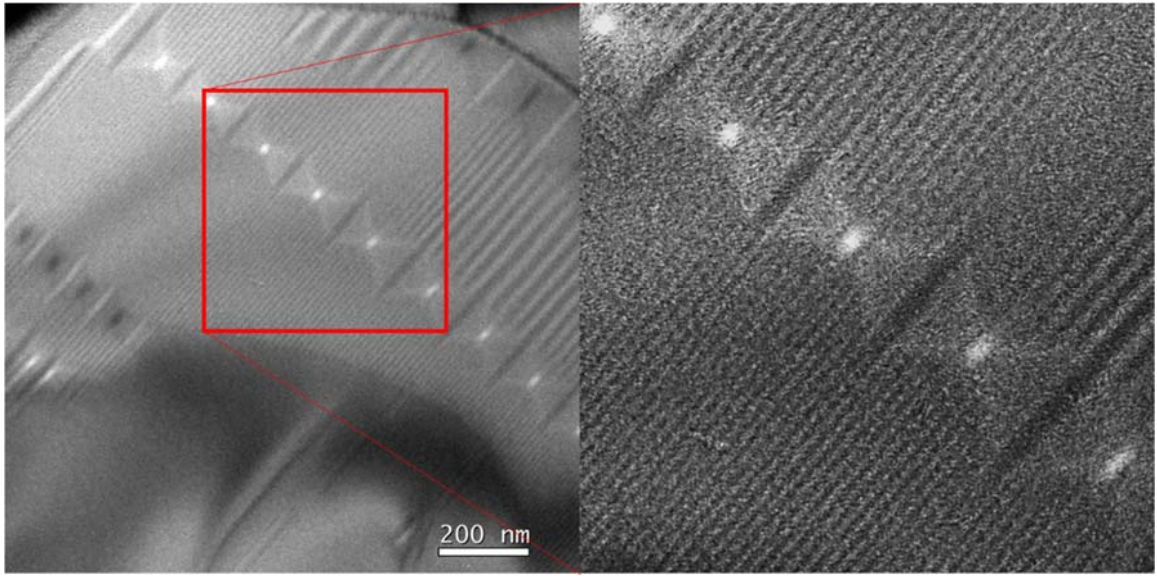

Fig. s5 Sandglass strings with straight stripes

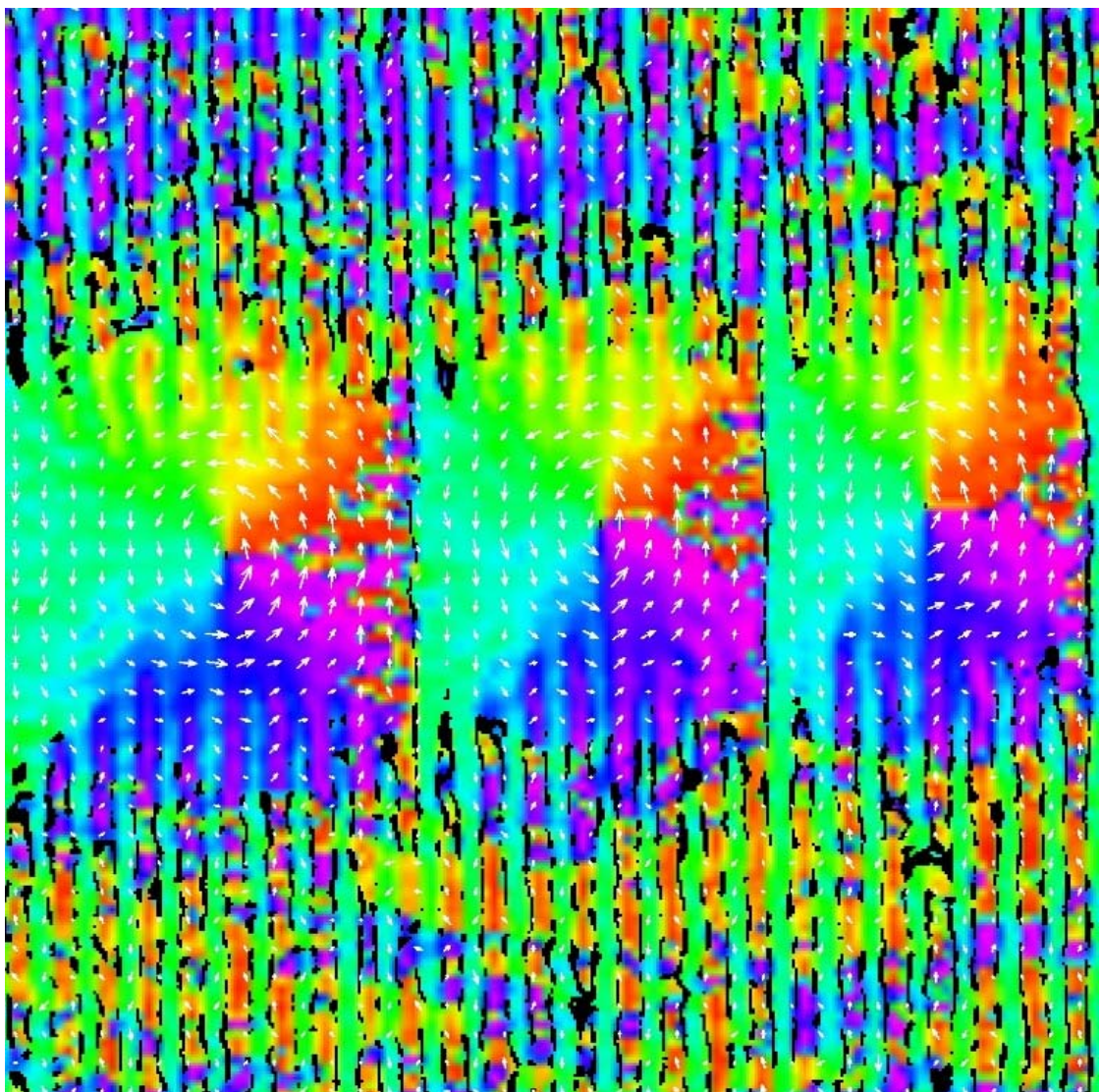

Fig. s6 The in-plane moment orientation mapping in Fig.4c

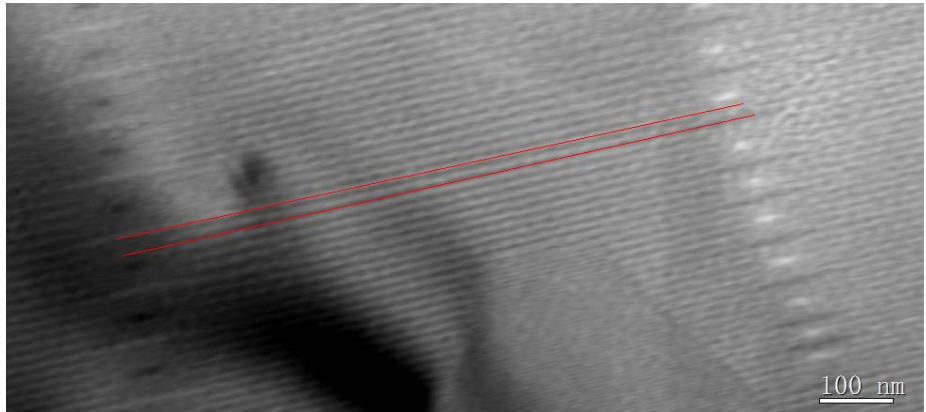

Fig. s7 The couple between two adjacent domain walls.

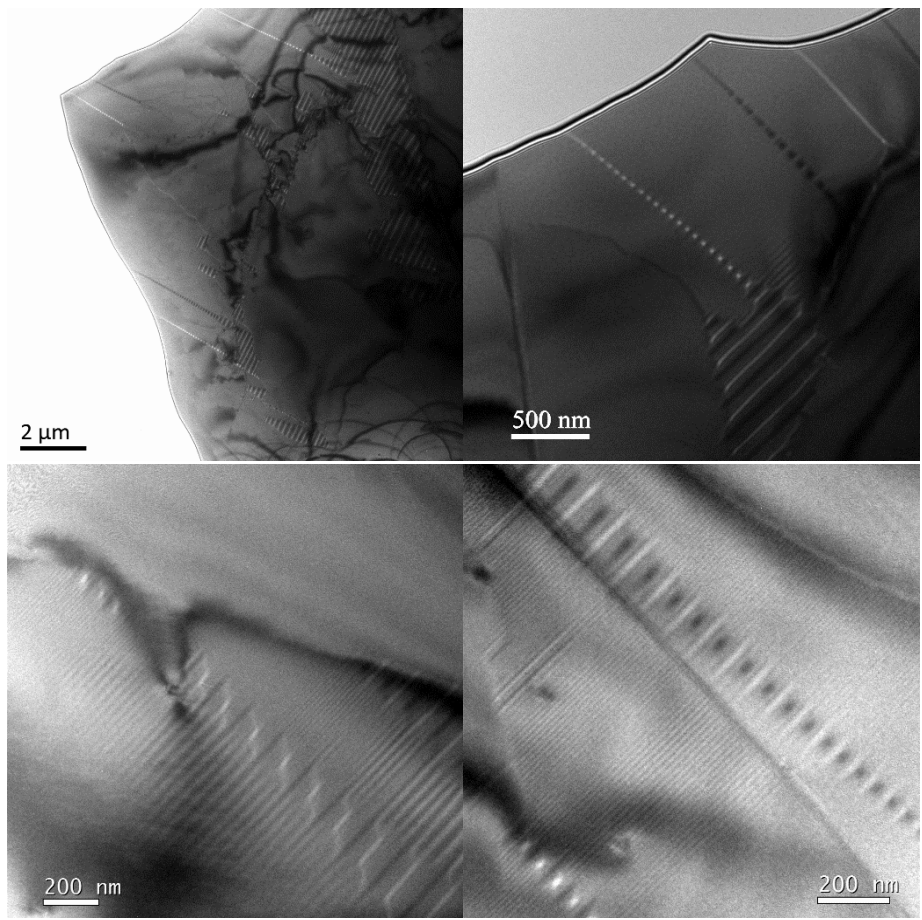

Fig. s8 The evolvement of the special domain walls at different regions.

Movie1: 3D distribution of Bx and By components
Movie2: Streamlines of the reconstructed B vectors in 3D space, colorized by the in-plane orientation decided by Bx and By components.